\documentstyle[prl,aps,epsf]{revtex}
\begin{document}
\preprint{Submitted to {\sl Physical Review Letters}}

\tighten
\draft
\twocolumn

\title{Quantum Logic Gates in Optical Lattices}
\author{Gavin K. Brennen,$^{(1)}$ Carlton M. Caves,$^{(1)}$
Poul S. Jessen,$^{(2)}$
and Ivan H. Deutsch$^{(1)}$}
\address{$^{(1)}$Center for Advanced Studies, Department of Physics and
Astronomy,\\
University of New Mexico, Albuquerque, NM 87131}
\address{$^{(2)}$Optical Sciences Center, University of Arizona,
Tucson, AZ 85721}
\date{\today}
\maketitle

\begin{abstract}
We propose a new system for implementing quantum logic gates: neutral atoms
trapped in a very far-off-resonance optical lattice. Pairs of atoms are
made to occupy the same well by varying the polarization of the trapping
lasers, and then a near-resonant electric dipole is induced by an auxiliary
laser. A controlled-NOT can be implemented by conditioning the target
atomic resonance on a resolvable level shift induced by the control atom.
Atoms interact only during logical operations, thereby suppressing
decoherence.
\end{abstract}


\tighten
\preprint{Submitted to {\sl Physical Review Letters}}

Any computation is constrained by the physical laws governing the machine
that carries out the operations. Conventional computers operate according
to the laws of classical physics, but an entirely new class of computers is
possible using physical components that are governed by the laws of quantum
mechanics \cite{Steane}. At the heart of quantum computation is the
entanglement of many two-state systems (qubits), which form the register of
the quantum computer. The requirements for creating and maintaining such a
highly entangled state seem to be almost contradictory: the qubits must be
strongly coupled to one another and to an external field to produce the
conditional-logic operations for quantum computation, yet coupling to other
external influences must be minimized because it leads to decoherence.
Quantum error correction \cite{Error} and fault-tolerant computation
\cite{FTol} promise to defeat the deleterious effects of decoherence, but
only if the coupling to the environment is sufficiently weak.

Several physical realizations of quantum computation have been proposed.  One of the 
most promising is based on storing each qubit in the state of an ultra-cold trapped 
ion \cite{Zoller}.  Ions interact strongly via their mutual Coulomb repulsion, thus allowing 
unitary manipulation of the qubits' joint state to be achieved with lasers \cite{Monroe}.  
Because of their charge, however, the ions interact strongly with the environment, giving 
rise to decoherence channels from technical noise sources \cite{Wineland};  possiblities for 
surmounting these problems are currently being explored \cite{IonHeat}.  Elements of 
quantum computation have also been implemented in standard NMR apparatuses 
\cite{NMR} and in cavity QED \cite{CQED}, but these schemes are at present 
difficult to scale to many qubits.  Solid-state systems, including quantum dots \cite{Qdot} 
have also been proposed for realizing quantum computation, but the strong interactions that 
exist in a condensed-matter environment make decoherence a difficult problem. A recent 
proposal \cite{Kane} to marry NMR techniques with silicon technology looks promising.

We propose here a new system for implementing quantum logic gates: trapped neutral 
atoms made to interact via laser-induced coherent electric dipole-dipole interactions.  Such a 
system has two advantages: decoherence is suppressed because neutrals couple weakly to 
the environment, and operations can be performed in parallel on a large ensemble of 
trapped atoms, thus offering avenues for scaling to many qubits. The main source of 
decoherence is spontaneous emission, but this can be negligible if all manipulations are 
performed rapidly compared to the photon scattering rate. To see that this is possible, 
consider the following scaling argument. The photon scattering rate is 
$\Gamma^{\prime}=s\Gamma /2 $, where $s$ is the saturation parameter, proportional to 
the excited state population, and $\Gamma \sim k^{3}\left| d_{eg}\right| ^{2}/\hbar$ is the 
spontaneous emission rate, $k$ being the wave number of the photon and $d_{eg}$ the 
dipole matrix element between the ground and excited states. For atoms spaced at distances 
small compared to the optical wavelength, retardation effects are negligible, and the level 
shift arising from the near-field dipole-dipole interaction scales as $V_{dd}\sim \left\langle 
d_{1}\right\rangle \left\langle d_{2}\right\rangle /r_{12}^{3}$, where $\left\langle 
d\right\rangle$ is the dipole expectation value and $r_{12}$  is the characteristic separation 
between the dipoles. For weak (nonsaturated) excitation $\left\langle d\right\rangle \sim 
\sqrt{s} \,d_{eg}$, so the ratio of interaction energy to scattering rate scales 
as $\kappa \sim ~V_{dd}/\hbar \Gamma ^{\prime }\sim \left(kr_{12}\right) ^{-3}$. Thus, 
if the atoms can be tightly confined to relative distances small compared to the wavelength, 
one can induce a coherent dipole-dipole interaction with negligible photon scattering. The 
central point is that the coherent level shift can be enhanced substantially through tight 
confinement, while the cooperative spontaneous emission rate cannot increase by more 
than a factor of two (the Dicke superradiant state) over that of an isolated atom. In addition, 
since the resonant dipoles can be turned ``on'' and ``off'' at will, atoms can be made to 
interact only during the conditional logic operations and not during single-qubit 
manipulations or during periods of free evolution, thereby reducing coupling to the 
environment.

As a concrete implementation, we consider here the use of neutral alkali atoms trapped in a 
far-off-resonance optical lattice, periodic potentials created  by a set of interfering laser 
beams in which atoms are trapped via the ac-Stark shift \cite{Opt-Lat}. By detuning the 
lasers very far from resonance, photon scattering is greatly reduced. Through a 
combination of near-resonance Sisyphus laser cooling and resolved-sideband Raman 
cooling \cite{Hamann}, atoms can be prepared in the ground state of the potential wells. In 
a recent experiment, $\sim 10^{6}$ Cs atoms were cooled in a two-dimensional optical 
lattice, with mean vibrational excitation $\bar{n}\approx 
0.01$ \cite{Hamann}. Such atoms can be tightly confined, with an rms spread on the order 
of $\Delta x\approx \lambda /50$ for reasonably deep wells, and are thus good candidates 
for inducing coherent dipole-dipole interactions. According to the discussion above, the 
ratio of the level shift to the linewidth is $\kappa =C/\eta ^{3}\approx 500\,C$, where $\eta 
=k\Delta x$ is the Lamb-Dicke parameter and $C$ is a number depending on the details of 
the geometry, to be determined below.

Consider a 3D optical lattice, detuned far to the blue of atomic resonance, with 
atoms trapped at the nodes thereby minimizing photon scattering, and which traps atoms 
deep in the Lamb-Dicke regime (see Fig.~1). The ``transverse'' beams confine the atoms in 
tubes oriented along the $z$ direction and the ``longitudinal'' beams produce standing 
waves of $\sigma _{+}$ and $\sigma_{-}$ light within each tube. The polarization 
gradient of the longitudinal fields allows the distinction of two ``types'' of atoms: those 
trapped at the nodes of $\sigma _{+}$-polarized wells and those trapped at nodes of 
$\sigma _{-}$-polarized wells. Central to our method is the ability to vary the lattice 
geometry dynamically: changing the angle $\theta$ between the longitudinal lasers' 
polarizations varies the distance $\delta Z$ between the minima of these wells according to 
$k_{L}\delta Z=\tan ^{-1}(\tan \theta/2)$. Two atoms trapped in neighboring wells can be 
brought into the same linearly polarized well by rotating the lasers' polarizations to parallel, 
adiabatically compared with the oscillation frequency in the well.  For large detunings 
stimulated Raman transitions by the lattice lasers are suppressed and the atom maintains its 
internal state.  Once in the same well, the atoms can be made to interact by applying an 
auxiliary ``catalysis laser'' that excites the atomic dipoles for a short time. A rotation of the 
laser polarization beyond $90^{\circ }$ slides the potential wells by more than a quarter 
wavelength. A given $\sigma_{+}$ atom can thus interact sequentially with all other 
$\sigma_{-}$ atoms, and an arbitrary entangled state within a tube can be created.

For each of the atomic types $\sigma _{\pm }$, we define a computational basis, $\left| 
1\right\rangle _{\pm },\,\left| 0\right\rangle _{\pm }$, of logical one and zero (see Fig.~2),
\begin{eqnarray*}
\left| 1\right\rangle _{\pm } &\equiv &\left| F_{\uparrow },M_{F}=\pm
1\right\rangle \otimes \left| n=0\right\rangle , \\
\left| 0\right\rangle _{\pm } &\equiv &\left| F_{\downarrow },M_{F}=\mp
1\right\rangle \otimes \left| n=0\right\rangle ,
\end{eqnarray*}
where $F_{\uparrow ,\downarrow }=I\pm 1/2$ denote the two hyperfine levels 
associated with the $S_{1/2}$ ground state and nuclear spin $I$ (half-integer), $M_{F}$ 
is the magnetic sublevel, and $\left| n=0\right\rangle $ is the vibrational ground state of the 
associated potential. Single-qubit operations can be performed via  pulses that are 
Raman-resonant with one type of atom. Two-qubit operations involve conditioning the 
state of one atom on the state of the other. For example, a C-NOT can be performed in the 
following way. Two atoms are made to reside in the same well as described above, and a 
weak $\pi$-polarized catalysis-laser field, propagating in the $x$-$y$ plane, is used to 
excite a near-resonant atomic dipole. If this laser is tuned to the 
$\left| S_{1/2},F_{\uparrow }\right\rangle \rightarrow 
\left| P_{3/2},F_{\rm\max}^{\prime }\right\rangle $ resonance, 
where $F_{\rm\max}^{\prime }=I+3/2$, a dipole is excited only if the atom is in the $\left| 
F_{\uparrow }\right\rangle $ state (i.e., the logical states $\left| 1\right\rangle _{\pm }$). 
The dipole-dipole interaction thus causes a shift only of the $\left| 1\right\rangle _{-}\otimes 
\left| 1\right\rangle _{+}$ two-qubit state and has neither diagonal nor off-diagonal matrix 
elements between any of the other two-qubit basis states. If the $\sigma _{-}$ type acts as 
the control bit and the $\sigma_{+}$ acts as the target, a Raman $\pi $-pulse on the shifted 
$\left| 1\right\rangle_{-}\otimes \left| 1\right\rangle _{+}\leftrightarrow \left| 
1\right\rangle_{-}\otimes \left| 0\right\rangle _{+}$ transition achieves a C-NOT with the 
usual truth table. The polarizations of the Raman lasers and an external magnetic field 
ensure that the pulse does not drive any other transition. Once the logic operation is 
executed the state of a register can be read out as in the ion trap \cite{Monroe}, by applying 
first a sequence of Raman pulses to isolate the population of that register in the $\left| 
F_{\uparrow}\right\rangle $ hyperfine state, and then detecting the amount of fluorescence 
on the cycling transition $\left| S_{1/2},F_{\uparrow }\right\rangle\rightarrow \left| 
P_{3/2},F_{\rm\max}^{\prime }\right\rangle$.

The dipole-dipole interaction is dependent both on the internal electronic states of the 
atoms, which determine the tensor nature of the interaction, and on the motional 
states, which determine the atomic wave-function overlap with the dipole-dipole potential. 
In the low saturation limit, the excited states can be adiabatically eliminated and the 
dipole-dipole interaction Hamiltonian between a pair of atoms can be written as 
$H_{dd}=V_{dd}-i \hbar \Gamma _{dd}$, where $V_{dd}$ describes the level shift and 
$\Gamma _{dd}$ describes the enhancement of the spontaneous photon scattering rate due 
to cooperative effects. For dipoles induced by the $\pi $-polarized catalysis laser, 
$H_{dd}$ is diagonal in the computational subspace, with the only nonvanishing matrix 
element given by \cite{Dipole-dipole}
\[
\left\langle 1_{+},1_{-}\right| H_{dd}\left| 1_{+},1_{-}\right\rangle =-\hbar \Gamma 
^{\prime }\,c_{g}^{4}\left\langle f\left( r,\theta_{r}\right) +ig(r,\theta _{r})\right\rangle ,
\]
where $c_{g}$ is the Clebsch-Gordan coefficient for the transition
$\left| F_{\uparrow },M_{F}=\pm 1\right\rangle \rightarrow \left|
F_{\rm\max}^{\prime },M_{F^{\prime }}=\,\pm 1\right\rangle.$ The functions $f$ and 
$g$ describe the dependence of the dipole-dipole interaction on the relative position of the 
two atoms:  $f+ig=ih_{0}^{(2)}\left(k r\right) +P_{2}\left( \cos \theta _{r}\right) 
\,ih_{2}^{(2)}\left(k r\right)$, where $h_{k}$ are spherical Hankel functions of order 
$k$, and $P_{2}(\mu )$ is the second-order Legendre polynomial. For small distances, 
$f$ scales as $1/r^{3}$, whereas $g$ goes to unity, corresponding to the full cooperativity 
of the superradiant state. The figure of merit is then given by $\kappa \equiv \langle 
V_{dd}\rangle /\langle \hbar (c_{g}^{4}\,\Gamma ^{\prime }+\Gamma_{dd})\rangle =-
\langle f(r,\theta_{r})\rangle /(1+\langle g(r,\theta_{r})\rangle ).$

Though spherically symmetric wells maximize the radial wave-function
overlap for atoms in their ground vibrational states, an isotropic relative
coordinate wave function is orthogonal to the $Y_{2}^{0}$ dipole potential.
We thus consider an axially symmetric harmonic potential with two atoms in
the vibrational ground state, each described by an elliptical Gaussian wave
packet with rms widths $ x_{0} \neq z_{0}$. Figure~3 shows a plot of $\kappa $, 
calculated numerically, as a
function $\eta_{\bot}=kx_{0}$ and $\eta_{\parallel}=kz_{0}$. Over the range
of values shown, $\left\langle g_{00}(r,\theta _{r})\right\rangle \approx
1$, i.e., full cooperativity. Given experimentally accessible localizations
$x_{0}=\lambda /60$ and $z_{0}=\lambda /30$, corresponding to
$\eta_{\bot}\approx 0.1$ and $\eta_{\parallel}\approx 0.2$, the figure of
merit is $\kappa \approx -19.3$. This is sufficient to resolve the level
shift and perform a two-bit logic gate with reasonable fidelity. An
approximate expression for $\kappa $, neglecting retardation effects, is
\[
\kappa \approx \frac{1}{8\sqrt{\pi }\eta_{\bot}^{2}\eta_{\parallel}}\!
\left[ -2-3%
\frac{\eta ^{2}}{\eta_{\bot}^{2}}+3\!\left( \frac{\eta ^{3}}{\eta_{\bot}^{3}}+%
\frac{\eta }{\eta_{\bot}}\right) \!\tan ^{\!-1}\!\left(
\frac{\eta_{\bot}}{\eta }
\right) \right] ,\,\,
\]
where $\eta ^{-2}\equiv \eta_{\parallel}^{-2}-\eta_{\bot}^{-2}$. Keeping
$\eta_{\bot}$ fixed and maximizing with respect to the ratio
$\eta_{\parallel}/ \eta_{\bot}$ gives $\kappa _{\rm\max}\approx
-0.017/\eta_{\bot}^{3}$ for a ratio $\left(
\eta_{\parallel}/\eta_{\bot}\right)_{\rm\max}\approx 2.18$. The small
prefactor stems mainly from the fact the relative coordinate rms in 3D is
at least $\sqrt{6}$ times the rms for a single particle in 1D.

An experiment to characterize the performance of the quantum gate can be performed as 
follows.  Consider the measurement of the C-NOT truth-table on an {\em ensemble} of 
identical atomic pairs.  After the atoms are cooled and the control and target atoms are 
prepared in one of the logical basis states, the gate operation is executed, and the four 
populations of the logical basis can be read out as described above. We can expect the 
occupation of atoms in an optical lattice to be random.  Atoms without the appropriate 
neighbor are in principle unaffected by the two-bit gate operation but generally will 
contribute to measured populations and reduce the apparent fidelity of the gate.  The 
background signal from unpaired atoms can be removed if we carry out additional 
measurements on a sample without any paired atoms, as well as measurements where the 
gate is operated once, population flushed from a register, and the gate operated a second 
time before readout.

As a concrete example, consider Cs atoms with lasers trapping blue of the 
$^{6}S_{1/2}\rightarrow\mbox{}^6P_{3/2}$ transition at $\lambda=852\,$nm. Given 
intensities of 100 W/cm$^{2}$ for all beams in Fig.~1 and detunings of 
$\Delta_{\bot}/2\pi \approx 120\,$GHz and $\Delta_{\parallel}/2\pi \approx 2\,$THz for 
the transverse and longitudinal lasers, respectively, we achieve the requisite localizations, 
$\eta_{\bot}=0.1$, $\eta_{\parallel}=0.2$ discussed above, corresponding to oscillation 
frequencies $\nu_{{\rm osc},\bot }\approx 200\,$kHz and $\nu _{{\rm 
osc},\parallel}\approx 50\,$kHz. For these parameters, photon scattering from the lattice 
produces a homogeneous linewidth $\Gamma_{{\rm lat}}^{\prime }/2\pi \approx 4\,$Hz.  
The catalysis laser is chosen as a perturbation to the trapping potentials, e.g., 
$\left|V_{dd}\right|/h=5\,$kHz.  Taking this field on atomic resonance and given a figure 
of merit $\kappa =\left| V_{dd}\right|/\Gamma _{{\rm cat}}^{\prime }=19.3$, the required 
catalysis intensity is $I_{{\rm cat}}\approx 0.1 \, \mu$W/cm$^{2}$, corresponding to a 
superradiant scattering linewidth $\Gamma_{{\rm cat}}^{\prime }/2\pi \approx250\,$Hz. 
Though the level shift depends sensitively on the atomic localization, it is a relatively weak 
function of the laser intensity since $\eta \sim I^{1/4}.$  These parameters 
should allow one to carry out a few logical operations with high fidelity.

Though optical lattices hold promise for producing entangled states of a few atoms, much 
remains to be done to implement even a rudimentary quantum computation. From the 
experimental side, one must develop a method for addressing and reading out the state of 
individual qubits, which are generally spaced closer than the optical wavelength. One 
possibility is to tag the atomic resonance to the position of a well by use of a gradient 
magnetic field or an additional ac-Stark shift as demonstrated in \cite{Gardner}. 
Alternatively, lattices could be designed with more widely separated wells through the 
superposition of many different wave vectors or the use of very long wavelength lasers 
(such as an intense CO$_{2}$ laser \cite{CO2}).  A major theoretical issue is the effect of 
atomic collisions \cite{Julienne}. For small interatomic distances the intense blue-detuned 
lattice fields will be resonant with a molecular potential which may lead to inelastic 
processes. Even for atoms in the electronic ground state, long-range molecular potentials 
can play an important role in atom-atom interactions for densities corresponding to two 
atoms in the same well. Though such interactions might be destructive for inelastic 
hyperfine-changing collisions \cite{Etite1}, elastic collisions might provide an alternative 
coherent coupling scheme with a lower decoherence rate \cite{Zoller2}, especially in the 
region of a Feschbach resonance that can be tuned with a magnetic field \cite{Etite2}. 
Another important question is that of error correction. For example, in our scheme each 1D 
``tube'' of atoms constitutes a separate quantum register acting in parallel with all the 
others. One might capitalize on this massive parallelism to increase the error threshold for 
fault-tolerant computation. Optical lattices are extremely flexible, with many experimental 
``knobs'', allowing a wide variety of possible mechanisms for implementing the essential 
features of quantum logic.

We thank Paul Alsing and John Grondalski for useful discussions. This work
was supported in part by New Mexico Universities Collaborative Research
(Grant No.\ 9769), the Office of Naval Research (Grant No.\
N00014-93-1-0116), the National Science Foundation (Grant No.\ PHY-9503259),
the Army Research Office (Grant No.\ DAAG559710165), and the Joint Services
Optics Program (Grant No.\ DAAG559710116).

\begin{figure}[tbp]
\caption{Schematic of a 3D blue detuned optical lattice.  Two pairs of $\pi$-polarized
beams $k_{\bot}$ provide transverse confinement, and the beams $k_{\parallel}$ (at a 
different frequency) provide longitudinal confinement in $\sigma _{+}$ and $\sigma _{-}$ 
standing waves. The solid (dotted) contours represent the resulting ellipsoidal potential 
wells associated with $\sigma _{+}$ ($\sigma _{-}$) polarization, separated pairwise by 
$\delta Z$, as a function of the relative polarization angle $\theta $.}
\end{figure}

\begin{figure}[tbp]
\caption{Schematic energy levels for the D2 line of a generic alkali in the
presence of a small longitudinal magnetic field (not to scale). The
computational basis states for atoms that follow $\sigma _{\pm}$ light are
indicated. The catalysis laser $\omega_{c}$ is near resonant for $\left|
1\right\rangle _{\pm }$ states. The transverse and longitudinal trapping
frequencies $\omega_{\bot}$ and $\omega _{\parallel}$ are detuned very far
to the blue of resonance. Unitary manipulation via Raman pulses connecting
only the $\left| 0\right\rangle _{+}$ and $\left| 1\right\rangle_{+}$
states is shown.}
\end{figure}

\begin{figure}[tbp]
\caption{Plot of $\kappa$, the ratio of the coherent dipole-dipole level
shift to the total linewidth, as a function of the Lamb-Dicke localization
parameter in the transverse direction, $\eta_{\bot}$, and the longitudinal
direction, $\eta_{\parallel}$.}
\end{figure}

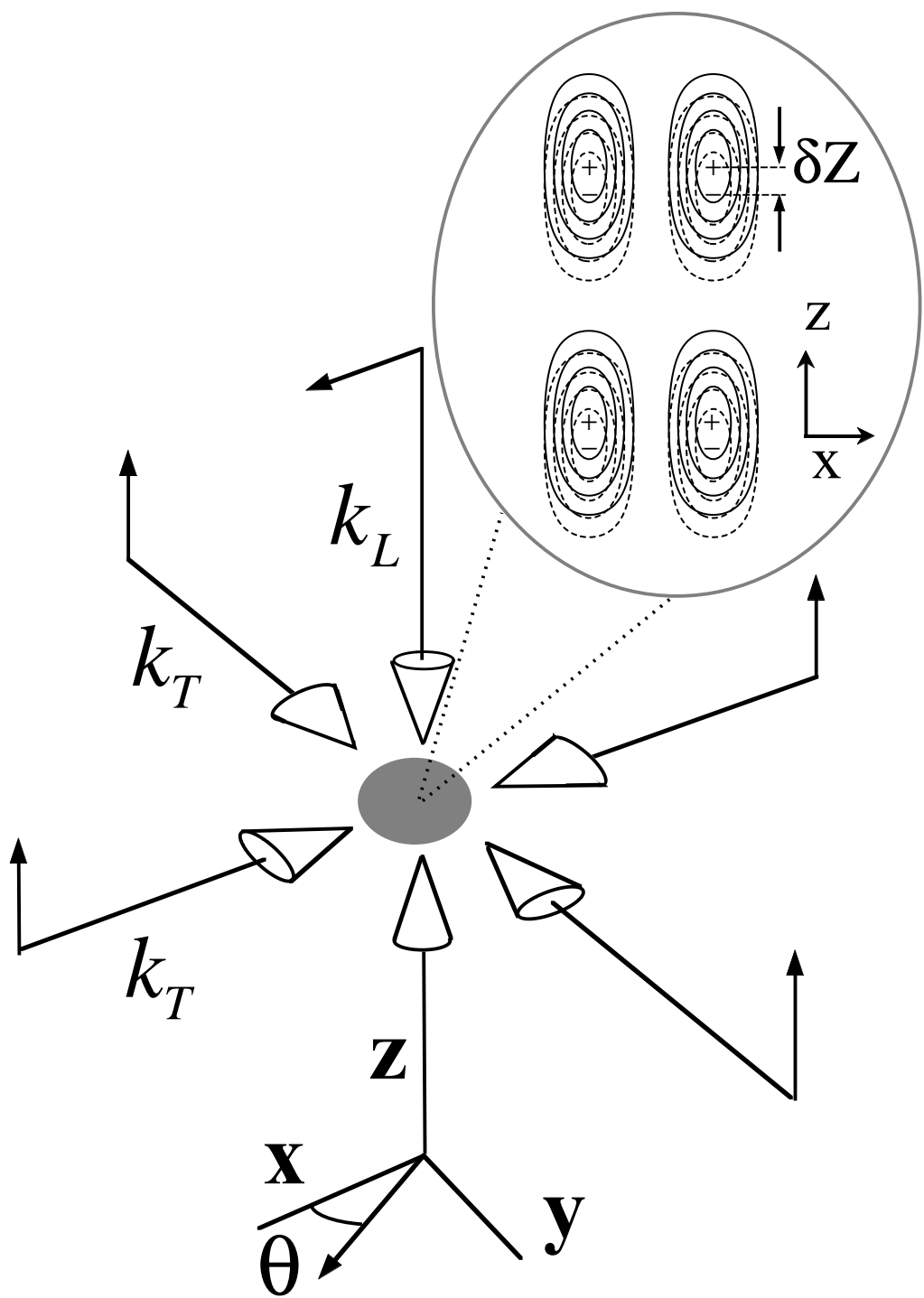
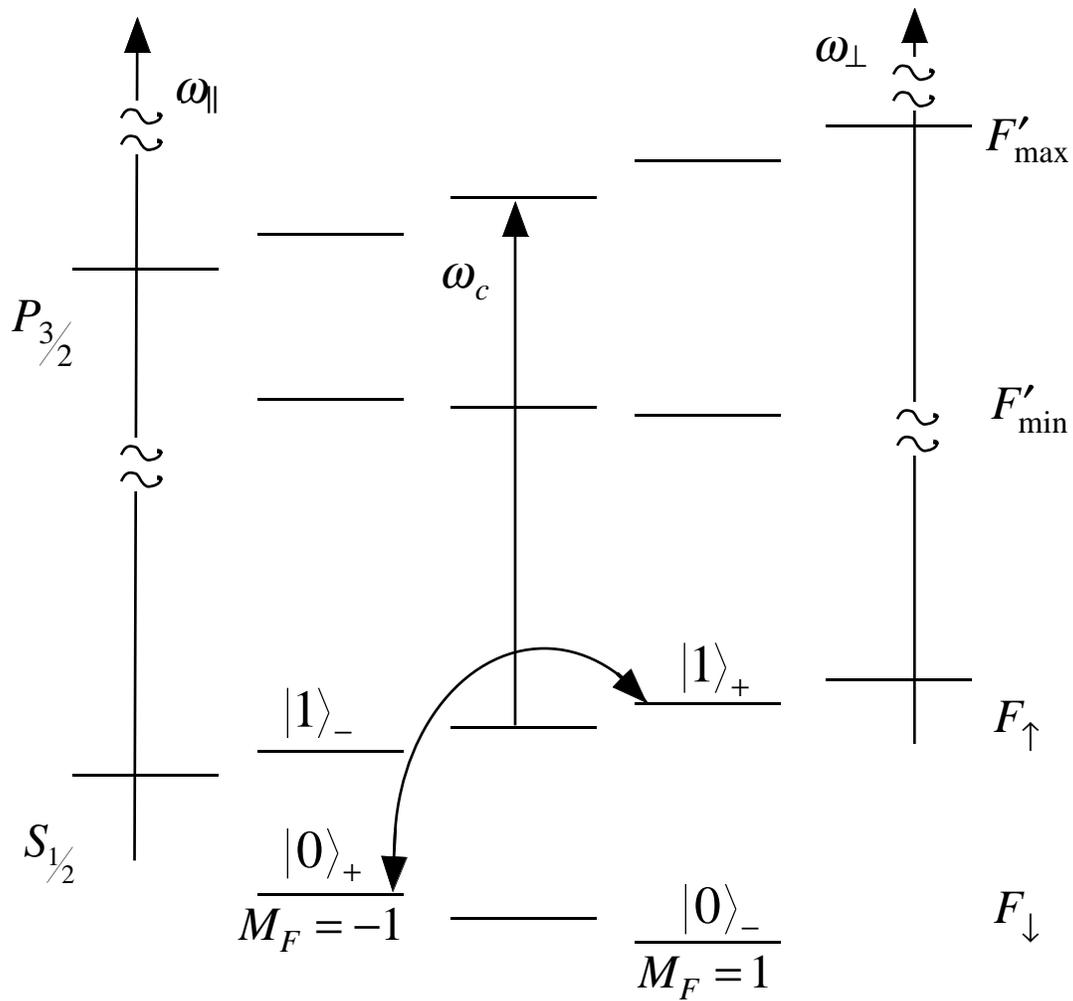
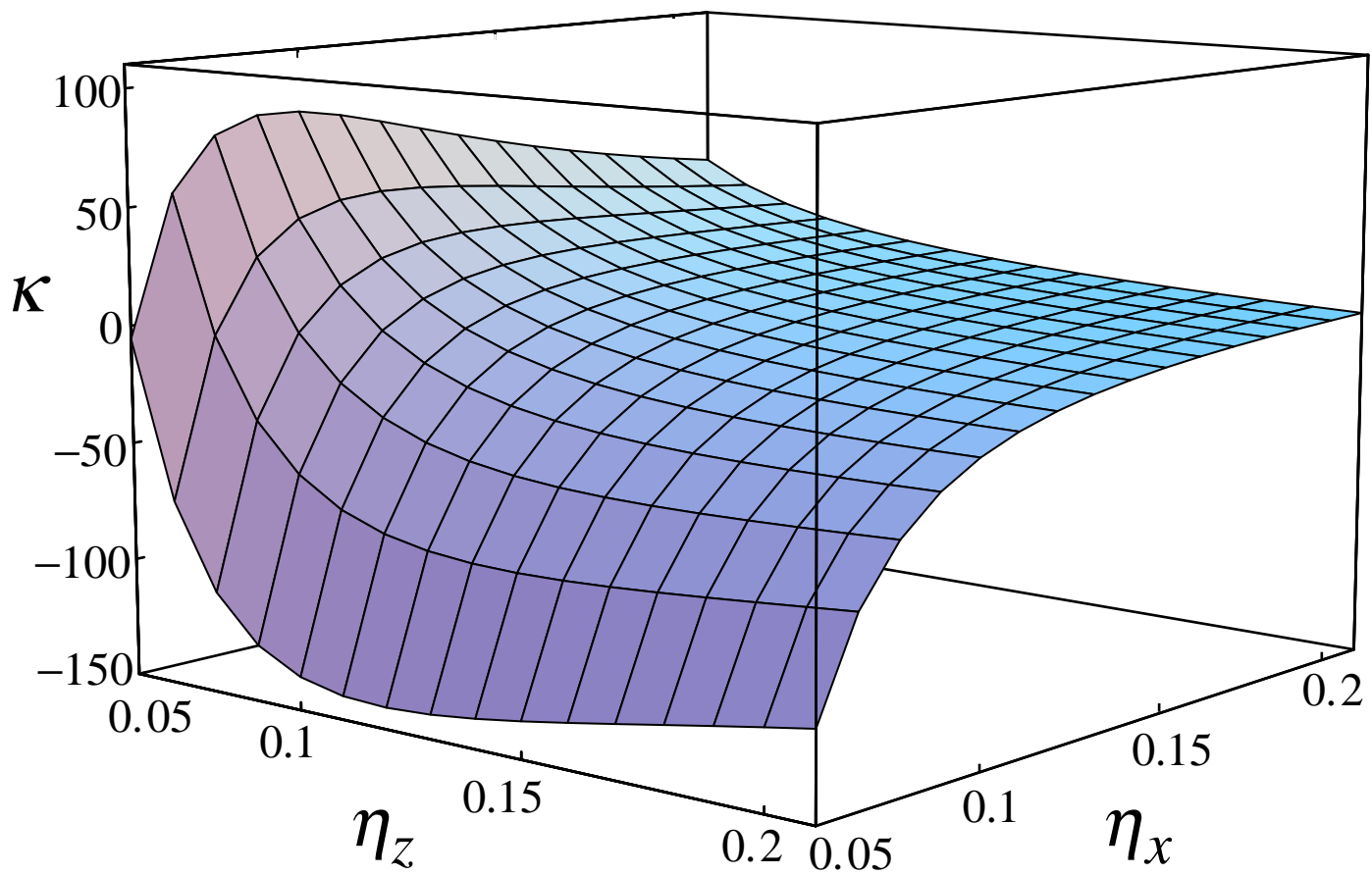
\end{document}